\documentclass[sigconf]{acmart}
\usepackage{url}

\usepackage{algorithmic}
\usepackage{graphicx}
\usepackage{textcomp}
\usepackage{float}

\usepackage{xcolor}
\def\BibTeX{{\rm B\kern-.05em{\sc i\kern-.025em b}\kern-.08em
    T\kern-.1667em\lower.7ex\hbox{E}\kern-.125emX}}
    
\setcopyright{acmlicensed}
\copyrightyear{2025}
\acmYear{2025}
\acmDOI{XXXXXXX.XXXXXXX} 

\begin{document}

\title{Cascaded Vulnerability Attacks in Software Supply Chains}

\author{Laura Baird}
\email{lbaird@uccs.edu}
\orcid{0009-0002-0538-7366}
\affiliation{
  \institution{University of Colorado Colorado Springs (UCCS)}
  \state{Colorado}
  \country{USA}
}

\author{Armin Moin}
\email{amoin@uccs.edu}
\orcid{0000-0002-8484-7836}
\affiliation{
  \institution{University of Colorado Colorado Springs (UCCS)}
  \state{Colorado}
  \country{USA}
}

\renewcommand{\shortauthors}{Baird and Moin}

\begin{abstract}
Most of the current software security analysis tools assess vulnerabilities in isolation. However, sophisticated software supply chain security threats often stem from cascaded vulnerability and security weakness chains that span dependent components. Moreover, although the adoption of Software Bills of Materials (SBOMs) has been accelerating, downstream vulnerability findings vary substantially across SBOM generators and analysis tools. We propose a novel approach to SBOM-driven security analysis methods and tools. We model vulnerability relationships over dependency structure rather than treating scanner outputs as independent records. We represent enriched SBOMs as heterogeneous graphs with nodes being the SBOM components and dependencies, the known software vulnerabilities, and the known software security weaknesses. We then train a Heterogeneous Graph Attention Network (HGAT) to predict whether a component is associated with at least one known vulnerability. Since documented multi-vulnerability chains are scarce, we model cascade discovery as a link prediction problem over CVE pairs using a multi-layer perceptron neural network. This way, we produce ranked candidate links that can be composed into multi-step paths. The HGAT component classifier achieves an Accuracy of 91.03\% and an F1-score of 74.02\%.
\end{abstract}


\maketitle

\section{Introduction}
Software Bills of Materials (SBOMs) are increasingly used to support software supply chain transparency and security management. For instance, the U.S. federal government has been mandating it for its contractors over the recent years \cite{TheWhiteHouse+2021_ExecutiveOrderStrengthening}. However, practitioners report persistent challenges in creating, maintaining, and validating SBOM content in realistic environments \cite{Xia+2023_EmpiricalStudySoftware,Stalnaker+2024_BOMsAwayMinds}. Even when an SBOM exists, studies \cite{ODonoghue+2024_ImpactsSoftwareBill,Benedetti+2024_ImpactSBOMGenerators} have shown that SBOM tool choices can affect vulnerability detection outcomes. Hence, the \textit{SBOM-in, vulnerabilities-out} pipelines can exhibit high variability across formats and tools. Moreover, SBOM generation and analysis tools must be enhanced to reduce the false-positive and false-negative rates in vulnerability detection, thus increasing efficiency \cite{Zhao+2024_CovSBOMEnhancingSoftware}. Typically, software security vulnerabilities are tracked using the Common Vulnerabilities and Exposures (CVE) records, with references to the Common Weakness Enumeration (CWE) entries, and prioritized using the Common Vulnerability Scoring System (CVSS). However, some software supply chain security exposure incidents demonstrate that multi-step combinations can turn individually manageable or low-severity issues into critical compromise paths. For example, the 2021 \textit{ProxyLogon} cyber-threat campaign chained multiple Microsoft Exchange vulnerabilities to allow an attacker to bypass the authentication system \cite{Microsoft+2021_HAFNIUMtargetingExchange}. Empirically, vulnerable dependencies are also frequently transitive and can persist for long periods before remediation, amplifying the practical risk of cascades across dependency chains \cite{Kumar+2024_ComprehensiveStudyImpact}. Recent research agendas in supply chain security increasingly emphasize the need to reason across components, ecosystems, and attacker techniques, not just severity scores \cite{Williams+2025_ResearchDirectionsSoftware}. The main contribution of this work is to propose a novel approach to co-exploitation path prediction across chains of security vulnerabilities and weaknesses within a software supply chain. The remainder of this extended abstract is structured as follows: Section \ref{sec:related-work} briefly reviews the literature. In Section \ref{sec:proposed-approach}, we propose our novel approach. Section \ref{sec:preliminary-results} reports our preliminary results. Finally, we conclude and suggest the future work in Section~\ref{sec:conclusion-future-work}.

\section{Related Work}\label{sec:related-work}

Yin et al. \cite{Yin+2024_CompactVulnerabilityKnowledge, Yin+2023_KnowledgeDrivenCybersecurityIntelligence, Yin+2023_EmpoweringVulnerabilityPrioritization} proposed novel approaches that are highly relevant to this work. They demonstrated that graph-based modeling significantly outperforms isolated vulnerability analysis by capturing the structural connections across disparate security databases. Their work focused on global vulnerability reasoning, whereas our proposed approach situates the relationships within the explicit dependency constraints of SBOMs, enabling the discovery of multi-step cascaded threats within software supply chains.


\section{Proposed Approach}\label{sec:proposed-approach}
We generate SBOMs in the \textit{CycloneDX} format using \textit{Syft} and enrich components with vulnerability information via \textit{Grype} backed by the \textit{Open Source Vulnerabilities (OSV)} database. We then convert each enriched SBOM into a heterogeneous graph with nodes for software components and dependencies, known security vulnerabilities (CVEs), and known security weaknesses (CWEs), linked by known relations among them. Besides heterogeneous nodes, our graph contains heterogeneous edges: \texttt{DEPENDS\_ON} between components, \texttt{HAS\_VULNERABILITY} from components to CVEs, and \texttt{HAS\_CWE} from CVEs to CWEs. Component features summarize scanner outputs and graph content (e.g., vulnerability count and CVSS aggregates, dependency metadata, and degree), CVE features encode severity and temporal metadata (e.g., CVSS-derived bins and recency), and CWE nodes use lightweight graph features (e.g., normalized degree). We then use a Heterogeneous Graph Attention Network (HGAT) backbone to learn over this heterogeneous structure, explicitly representing multiple node and edge types and using attention to surface which relations are informative for downstream tasks \cite{Yang+2021_HGATHeterogeneousGraph}. Specifically, we employ the multi-head graph attention method \cite{Velickovic+2018_GraphAttentionNetworks} with two heads and learned attention coefficients per edge type, allowing the model to weight the importance of different neighbors dynamically. The task is to predict whether a component is associated with at least one known vulnerability. Moreover, predicting full multi-CVE chains directly is currently challenging due to a lack of semantic, real-world chains of vulnerabilities for supply chain attacks. To make progress under limited supervision, we model cascade discovery as \textit{link prediction} over pairs of CVEs: given two vulnerabilities, estimate the likelihood that they will be co-exploited to facilitate an attack. We implement a lightweight feature-based Multi-Layer Perceptron (MLP) neural network predictor trained in a few-shot regime using domain-informed features derived from vulnerability metadata (e.g., severity, temporal proximity, weakness information, and exploitation signals), with negative sampling to construct non-chain pairs. The resulting ranked CVE links can be composed into candidate multi-step chains by chaining high-scoring edges.






\section{Preliminary Results}\label{sec:preliminary-results}
To build a small training seed set for cascade prediction, we extract CVE sequences from public advisories and incident reports that explicitly describe chained exploitation. We retain only within-chain CVE pairs as positives and sample non-chain pairs as negatives (2:1 negatives to positives). Using the Wild SBOMs dataset \cite{Soeiro+2025_WildSBOMsLargescale}, we train HGAT-based node classification to predict a coarse label (\texttt{has-any-CVE}) on a random subset of 200 Python-based CycloneDX SBOMs. On our current split, the model achieves 91.03\% Accuracy, 80.84\% Precision, 68.26\% Recall, and 74.02\% F1-score. Ablating dependency edges reduces positive predictions, suggesting that relational structure contributes a signal beyond local component metadata, supporting efficient downstream path ranking. Furthermore, we evaluate the cascade predictor on a seed set of documented multi-CVE chains (n=35 documented chains: 27 from vulnerability disclosures and 8 from incident reports) paired with sampled non-chain pairs. On this initial setting, the model achieves 0.93 ROC-AUC, with clear separation between chain and non-chain pairs. This evaluation is limited by chain scarcity and by pair-level splitting (the same CVE may appear in both train and test via different pairings); we are implementing chain-level and temporal splits to better measure generalization to unseen chains. The source code is available as open-source software at \url{https://github.com/qas-lab/laura-baird-icse2026}. All research data are publicly available as cited above and at \url{https://doi.org/10.7910/DVN/A6CZRB}.



\section{Conclusion and Future Work}\label{sec:conclusion-future-work}
We have proposed a novel approach to predicting cascaded vulnerabilities in software supply chains from SBOMs. In the future, we plan to produce semantic research data from public sources to extend our evaluation and enhance the current prototype using a variety of AI-based methods and techniques, such as Large Language Models (LLMs) and knowledge graphs.


\section*{Acknowledgment}
This material is based upon work supported by the U.S. National Science Foundation (NSF) under Grant No. 2349452. Any opinions, findings, conclusions, or recommendations expressed in this material are those of the authors and do not necessarily reflect the views of the NSF. We also used generative AI tools for content and programming.

\bibliographystyle{ACM-Reference-Format}
\bibliography{References}

\end{document}